\documentclass[12pt]{iopart}

\usepackage{iopams}  
\usepackage{graphicx}
\usepackage{subfigure}
\usepackage{citesort}
\usepackage{subfigure}
\usepackage[usenames]{color}

\begin{document}

\title[]{Overcoming the asymmetry of the electron and hole doping for magnetic transitions in bilayer CrI$_{3}$}

\author{Sukanya Ghosh*, Nata{\v s}a Stoji{\'c} and Nadia Binggeli}

\address{Abdus Salam International Centre for Theoretical Physics, Strada Costiera, Trieste, Italy}
\ead{*sghosh4@ictp.it}
\vspace{10pt}

\begin{abstract}
Electrical control of magnetism has great potential for low-power spintronics applications and the newly discovered two-dimensional van-der-Waals magnetic materials are promising systems for this type of applications. In fact, it has been recently shown experimentally (Jiang {$\textit{et al.}$} 2018 {$\textit{Nat. Nanotechnol.}$} {\textbf{13}}, 549–553) that upon electrostatic doping by electrons bilayer CrI$_3$ undergoes an  antiferromagnetic-ferromagnetic (AFM-FM) phase transition, even in the absence of magnetic field.  Doping  by  holes,  on the other hand, does  not  induce  the  same  transition in the experiment, which points  to  an  intrinsic asymmetry in the hole and electron doping that limits the  control of the transition by doping.  We here show, based on first-principles calculations, that the asymmetry originates in the relativistic nature of the valence-band-edge states of the pristine bilayer, which inhibits the magnetic transition upon hole doping. Based on this finding, we propose  an  approach  to engineer this system so that it displays the AFM-FM transition for both hole and electron doping by  using  moderate uniaxial strain along the soft direction of the bilayer. 

\end{abstract}

%

%
%
%
%
%

\section{Introduction}

Electrical control of magnetism is key for the development of future applications in spintronics and magnetic data storage, because of its low-power dissipation and high speed \cite{SONG201733,Matsukura2015}. The recently discovered two-dimensional (2D) magnetic materials \cite{GonLiLi17,HuangNature2017}, which are easily integrated and gated in van der Waals (vdW) heterostructures \cite{GibKop19,Tsymbal2019,Nardelli2020,Stroppa2020,DolPetZol20,GonZha19}, provide exceptional opportunities for manipulation and switching of magnetism in nanoscale devices \cite{GibKop19,Nanoscale,GonZha19}. For magnetic state control, the CrI$_3$ bilayer has been the focus of particular attention \cite{GibKop19,GonZha19}. CrI$_3$ is one of the first emerged and the most studied of the 2D magnetic semiconductors \cite{HuangNature2017,GibKop19}. Interestingly, its magnetic properties depend on the number of layers. In the bulk,  CrI$_3$ is a ferromagnetic (FM) semiconductor with Curie temperature $T_C=61$~K \cite{McGuire2015}. The CrI$_3$ monolayer has a FM ground state with $T_C=45$~K \cite{HuangNature2017}, while the bilayer is experimentally observed to be a layered antiferromagnet  \cite{HuangNature2017,Ubrig_2019,Thiel973,JiaShaMak18,HuaClaKle18} with N{\' e}el temperature similar to the monolayer  $T_C$  \cite{JiaShaMak18,HuaClaKle18}. 

The presence of relatively weak interlayer magnetic interactions in  CrI$_3$ and the number-of-layer dependent magnetic properties make them ideally tailorable by different approaches. In the bilayer, the small interlayer exchange coupling is evidenced by a low magnetic critical field of 0.6-0.7~T for the antiferromagnetic (AFM) to FM transition  \cite{HuangNature2017}. In recent experiments, this spin-flip field could be tuned by several techniques, including in-plane strain \cite{JiaXieSha20}, hydrostatic pressure \cite{Song2019,Li2019}, and electrostatic gating \cite{JiaShaMak18,HuaClaKle18}, while an irreversible AFM to FM transition was observed at high hydrostatic pressure (above 1.7 GPa) \cite{Song2019,Li2019}. Theoretically, the exchange coupling was also shown to be sensitive to perturbations such as strain \cite{LeoGonMej20}, change in layer stacking \cite{SivOkaXu18,JiaWanChe19,SorCarFer19}, electric field \cite{Su_rez_Morell_2019,XuZou20}, electrostatic doping \cite{JiaWanChe19,SorKat20,LeiChi19}, and by coupling to other 2D materials  \cite{LiuZhoZha20} -- in most cases with auspicious trends for FM transitions \cite{LeoGonMej20,SivOkaXu18,JiaWanChe19,SorCarFer19,Su_rez_Morell_2019,XuZou20,SorKat20,LeiChi19,LiuZhoZha20}. 
Each of the external perturbations 
acts through a different mechanism, offering much-needed versatility for diverse applications. Electrostatic doping modifies orbital occupation, while the electric field causes a layer-dependent 
energy shift of the bands, tending to favor ferromagnetism \cite{Su_rez_Morell_2019}. Hydrostatic pressure was observed  experimentally to induce an FM transition by changing the stacking order \cite{Li2019}, while in-plane tension, via inverse magnetostriction, was found to lower the spin-flip field  \cite{JiaXieSha20}.

Specific to electrical control, remarkable experimental manipulations of 2D magnetic states by electrostatic gating have been achieved recently in bilayer CrI$_3$  \cite{JiaShaMak18,HuaClaKle18,JiaLiWan18}. The use of dual-gate field-effect-device configurations enabled independent gating control over the electric field and electrostatic doping.  While the electric field was observed to induce, at zero magnetic field, an interesting linear magnetic-polarization effect in the magneto-optical Kerr signal \cite{JiaShaMak18,HuaClaKle18}, electron doping monotonously  decreased the spin-flip field, and correspondingly enabled, at fixed magnetic field near the critical field, a voltage-controlled switching between AFM and FM states \cite{HuaClaKle18,JiaLiWan18}. 

Strikingly, in the latest voltage-controlled doping experiments on graphene-encapsulated CrI$_3$ bilayer \cite{JiaLiWan18}, the spin-flip field was made to vanish above a critical electron density, and electron doping was demonstrated to induce a transition from interlayer antiferromagnetism to ferromagnetism, in the absence of a magnetic field. 
Upon hole doping, on the contrary, no transition was observed at zero field. The asymmetry of electron and hole doping for the magnetic transition is consistent with the trend of a recent ab-initio calculation for the isolated CrI$_3$ bilayer  \cite{JiaWanChe19},  confirming it is an intrinsic feature of the bilayer. 
The asymmetry or lack of magnetic transition with hole doping, however, is an intriguing feature, as it is contrary to the general expectation that weak doping favors ferromagnetism \cite{Aka98}, and also because some (but not all \cite{JiaWanChe19})  theoretical studies indicated a symmetric behavior \cite{SorKat20,LeiChi19}. The symmetry in the latter studies might be related to neglect of relativistic effects, but the factors controlling the asymmetry in the magnetic behavior have not been investigated so far. 
Furthermore, the existence of theoretical results indicating a symmetric behavior raises the interesting possibility that through band-structure engineering one may be able to lift the asymmetry in the experimental bilayer and control magnetism also with hole doping.  This would mean increased flexibility and enhanced magnetic switching ability, allowing electrical control of the AFM-FM phase transition by both positive and negative bias in the experiment. 

In this paper, using ab-initio density-functional-theory (DFT) calculations, we find the origin of the asymmetric behavior of the hole doping compared to electron doping.  Using this finding, we propose an approach to engineer the system mechanically, so that it undergoes an AFM-FM phase transition both with hole and electron doping by applying moderate compression perpendicular to the bilayer.

\section{\label{sec:level1} System and  computational method 
}

\begin{figure*}[t]
\begin{center}
\includegraphics[width=0.52\textwidth] {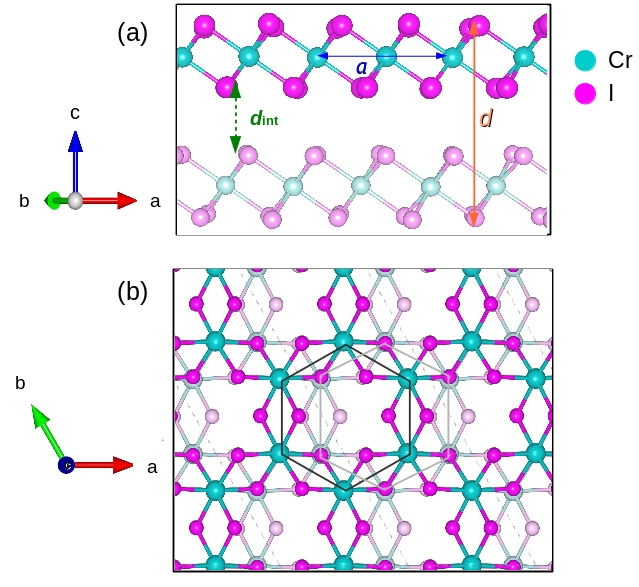}
\caption{\label{fig:structure}Side (a) and top (b) view of the atomic structure of the CrI$_{3}$ bilayer with monoclinic layer stacking. In a single CrI$_{3}$ layer, the plane of Cr atoms forms a honeycomb lattice and is sandwiched  between  two  I atomic  planes.  Cr  ions  are  surrounded  by six  first  neighbor  I ions  arranged  in corner  sharing  I octahedra; {\bf a}, {\bf b}, {\bf c} are the unit-cell vectors, and $d$ ($d_{int}$) is the distance between the outermost (innermost) iodine atomic planes of the bilayer.  }
\end{center}
\end{figure*}

We consider the CrI$_{3}$ bilayer with monoclinic layer stacking, as shown in figure~\ref{fig:structure}. This corresponds to the structural phase of the CrI$_{3}$ bilayers displaying experimentally layered antiferromagnetism at low temperature, and obtained by exfoliation from CrI$_3$ bulk van der Waals crystals at room temperature  \cite{HuangNature2017,Ubrig_2019,Thiel973,JiaShaMak18,HuaClaKle18,JiaXieSha20,Song2019,Li2019,JiaLiWan18}. 

Bulk CrI$_{3}$ exhibits a low-temperature (LT) structure with rhombohedral layer stacking and R$\overline{\rm 3}$ space-group symmetry, and a high-temperature (HT) structure with monoclinic layer stacking and C2/m symmetry, corresponding to temperatures below and above $\sim$200~K, respectively  \cite{McGuire2015}. 
CrI$_3$ bilayer samples obtained by exfoliation from the bulk at room temperature have been shown experimentally to remain in the monoclinic phase also at low temperature \cite{Ubrig_2019,Thiel973}. The monoclinic stacking gives  rise to a layered AFM ground state, while rhombohedral stacking yields an FM ground state  \cite{SivOkaXu18,JiaWanChe19}. 
 Hence, although in Ref.~\cite{JiaLiWan18} the experiment reporting  the AFM-FM transition upon electron doping was performed at low temperature, the bilayer CrI$_3$ samples used were  fabricated at room temperatures and then encapsulated and cooled, remaining in  the AFM monoclinic phase due to the existence of an energy barrier for the transition to the rhombohedral phase \cite{JiaLiWan18,JiaShaMak18,Song2019}

In our study, all structural and electronic properties were calculated using DFT in its plane-wave pseudopotential implementation  in the Quantum ESPRESSO package \cite{GiaBarBon09}. We performed calculations within the local-spin-density approximation (LSDA), with and also without spin-orbit coupling (SOC), by using fully-relativistic  projector augmented wave (PAW) pseudopotentials and comparing the results to those obtained with the scalar-relativistic version of the PAW pseudopotentials.  The plane-wave kinetic energy cutoffs for the electronic states and charge density were chosen to be 60 Ry and 650 Ry, respectively, both for the fully relativistic and scalar-relativistic cases. The Brillouin zone (BZ) was sampled using the $\Gamma$-centered uniform Monkhorst-Pack $k$-point grid of $24 \times 24 \times 1$. 

Our optimized in-plane lattice constant is $a = 6.69$~\AA\ and the equilibrium distance between the two outermost I atomic planes of the bilayer is $d^{\rm eq}=9.47$~\AA, while the separation between the two inner I atomic planes in the bilayer is $d_{\rm int}^{\rm eq}=~3.36$~\AA\ (see figure~\ref{fig:structure}). 
The lattice parameters obtained using the scalar-relativistic pseudopotentials remained virtually unchanged upon relaxation performed with the relativistic pseudopotentials. The periodic images of the CrI$_{3}$ bilayer were separated by a large vacuum distance of $\sim$30 \AA\ along the out-of-plane direction. Unless otherwise specified the relaxations were performed with the undoped systems. For the  structural relaxations, the atomic coordinates were relaxed until the forces on the atoms became less than 0.1~mRy/Bohr. 
For the electrostatic doping, we added electrons or holes to the bilayer system and used a compensating jellium  background in the supercell to maintain charge neutrality. 

For the strained bilayer, we applied strain along the out-of-plane direction by changing the distance $d$ between the two outermost iodine atomic planes of the bilayer. For a given uniaxial strain, the vertical coordinate of the I atoms in these surface planes were fixed to the distance imposed by the strain, while all other atomic coordinates were free to relax, and the in-plane lattice parameter was also  optimized. The change of the in-plane lattice parameter was rather small, it expanded by 1\% upon  a  vertical compression  $\epsilon_{zz} = - 5$\%. 

We checked the stability upon strain of the out-of-plane orientation of the Cr spin, by calculating the magnetic anisotropy energy of the pristine and strained bilayer. 
We found it to be 0.65~meV/Cr and 0.36~meV/Cr for $\epsilon_{zz} = 0$ and $-5$\%, respectively. Although reducing, the magnetic anisotropy energy did not change sign, showing the same preferred perpendicular spin orientation also in the strained case. In addition, we evaluated the effect of the 5\%-vertical compression on the energy barrier between the HT-monoclinic and LT-rhombohedral phases of the CrI$_3$ bilayer and found that  the energy barrier is increased with respect to the unstrained case  (see Appendix). This thus tends to further inhibit a spontaneous structural transition from the HT to LT phase.

It should be noted that the overall bilayer deformations (the aspect ratio, in particular) induced by the vertical strains applied along the soft direction, which we consider in this work, differ significantly from those induced by in-plane biaxial strains. In fact, we find that a tensile in-plane strain $\epsilon_{xx}=\epsilon_{yy}=$1\% on the bilayer would induce a $d$ contraction of only 0.5\%. This corresponds to  a ratio $\epsilon_{zz}/\epsilon_{xx}=-0.5$, which is 10 times smaller in magnitude than with the applied uniaxial perpendicular strains (see Appendix). We also note that compared to vertical electric fields \cite{GhoStoBin19,comment_SG}, the vertical strains have a non-negligible influence on the internal atomic structure of each CrI$_3$ monolayer (as described in the Appendix).

\section{\label{sec:level1}Results and Discussion} 
\subsection{\textbf{Origin of the asymmetry of electron and hole doping for the magnetic transition }}

\begin{figure*}[t]
\begin{center}
\includegraphics[width=0.45\textwidth] {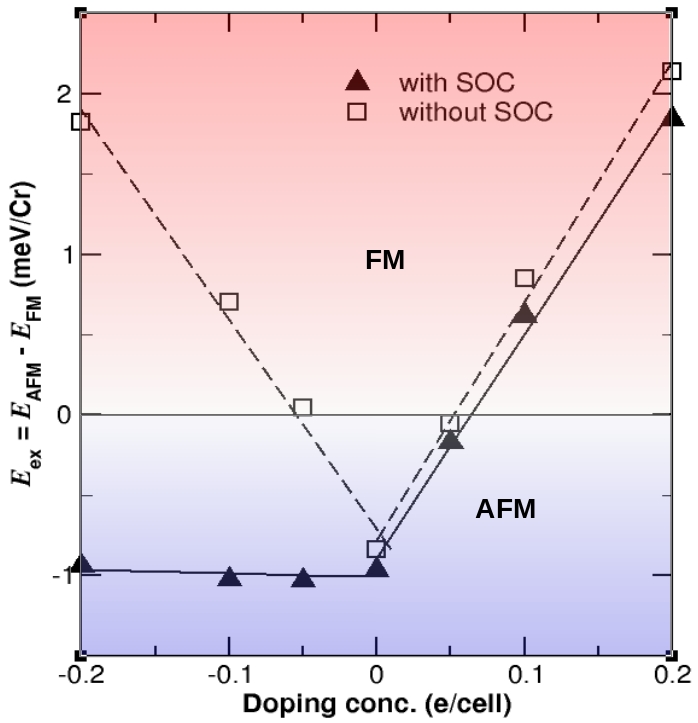}
\caption{\label{fig:Eex} Exchange energy, $E_\mathrm{ex}$, as a  function of doping for the CrI$_{3}$ bilayer from fully-relativistic and scalar-relativistic calculations. Electron doping is represented with positive and hole doping with negative sign.  }
\end{center}
\end{figure*}

In figure~\ref{fig:Eex} we show our calculated exchange energy, $E_\mathrm{ex} = E_\mathrm{AFM}- E_\mathrm{FM}$, as a function of electrostatic doping, from scalar-relativistic and fully-relativistic calculations. In both cases, at zero doping,   $E_\mathrm{ex}$ is negative (with similar values), consistent with the experimentally observed AFM ground state of the bilayer. 
In the scalar-relativistic case,  $E_\mathrm{ex}$ increases linearly with electron and hole doping, displaying a symmetric behavior with a transition from AFM to FM occurring both for electron and for hole doping (at about 0.052 electrons/cell and 0.054 holes/cell, respectively). 

With the inclusion of SOC, the hole doping does not induce the AFM-FM transition anymore up to large doping. In fact, $E_\mathrm{ex}$  remains virtually constant for hole doping up to at least 0.2 hole/cell, while with electron doping the exchange energy keeps increasing linearly, displaying about the same slope as in the scalar-relativistic case. The asymmetry in the magnetic behavior  induced by electron and hole doping corresponds fully to the experimental findings \cite{JiaLiWan18} -- yielding even the transition electron-doping value (1.7$\times 10^{13}$~e/cm$^2$ or 0.064 electron/cell) relatively close to the experimental one (around 2.0$\times 10^{13}$ e/cm$^2$), and the results in figure~\ref{fig:Eex} thus show that the asymmetry with lack of transition for hole doping has a relativistic origin.

\begin{figure*}[t]
\begin{center}
\includegraphics[width=0.55\textwidth] {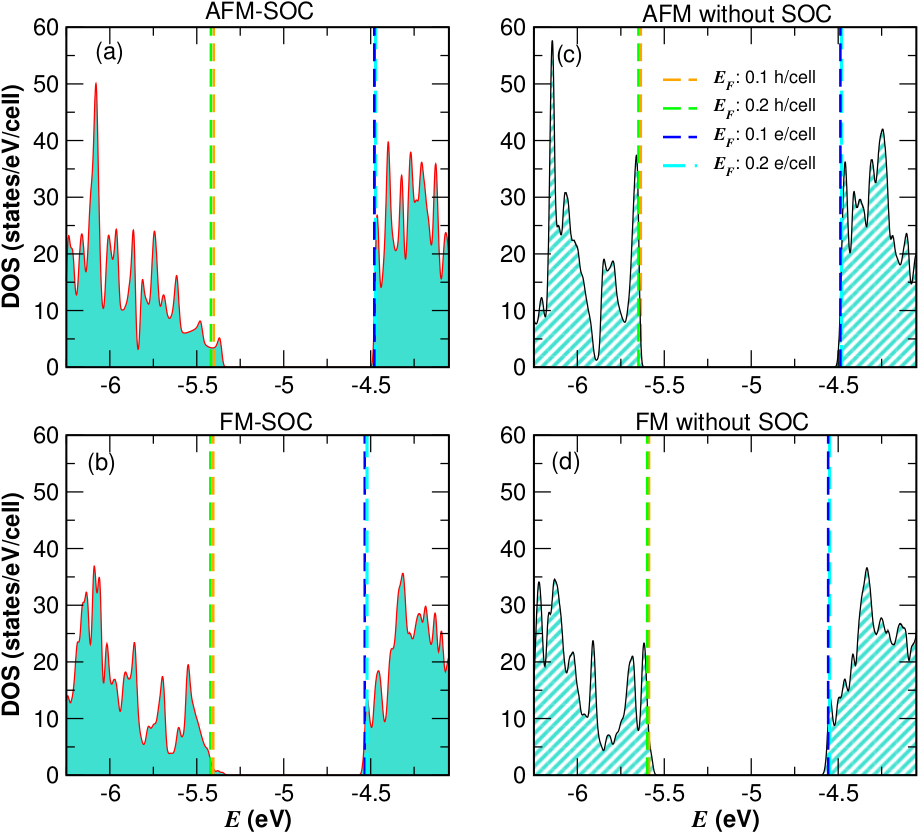}
\caption{\label{fig:DOS} Electronic density of states of the CrI$_{3}$ bilayer in AFM (a,c) and FM (b,d) configuration, as obtained in the fully-relativistic  (a-b)  and (c-d) scalar-relativistic calculations. The dashed lines indicate the position of the Fermi level obtained in the rigid-band model for 0.1, 0.2 hole/cell doping (in yellow, green, respectively) and 0.1, 0.2 electron/cell doping (blue, cyan, respectively). The energy scale is relative to the vacuum potential. }
\end{center}
\end{figure*}

In order to better understand the factors controlling the asymmetry and presence or absence of a magnetic transition with hole doping, we have examined the electronic density of states (DOS) and band structure of the bilayer in the fully-relativistic and scalar-relativistic cases. In figure~\ref{fig:DOS}, we show the DOS's obtained for the pristine AFM and FM bilayers in the fully-relativistic (a-b) and scalar-relativistic (c-d) calculations. In each case, the electronic energies are measured relative to the vacuum potential, allowing comparison between the band energies of the AFM and FM bilayer. In fact, as the electrostatic potentials for the AFM and FM bilayers are found to be virtually identical (see Supplementary Information), one may expect the change in $E_\mathrm{ex}$ upon weak electron/hole doping to be dominated by the difference in the AFM and FM one-electron energy contribution from the occupied/emptied band-edge levels in their respective conduction/valence DOS spectra. 

Inspection of figure~\ref{fig:DOS} indicates that, in both AFM and FM bilayers, the presence of SOC drastically influences the DOS near the valence band maximum (VBM) and significantly alters the VBM energy, while the conduction band minimum (CBM) is affected much less. The VBM in the scalar-relativistic case is characterized by a very sharp DOS feature with large DOS values, which is replaced in the fully relativistic case by a considerably smoother DOS with lower DOS values near the VBM.   

In figure~\ref{fig:DOS}, we also indicated the positions of the Fermi level, $\epsilon_F$, for 0.1, 0.2 electron/cell doping and for 0.1, 0.2 hole/cell doping, as obtained using the rigid-band model of the pristine bilayers (in good agreement with results for band structures and corresponding $\epsilon_F$ positions from self-consistent doping calculations, see Supplementary Information). In all cases, $\epsilon_F$ lies rather close to the band edge (within 0.1 eV).  

In the scalar-relativistic case (figure~\ref{fig:DOS}(c) and (d)), the FM phase has both conduction and valence DOS noticeably broader than the AFM phase; the FM phase has a VBM and $\epsilon_F$ for hole doping at higher energy, and a CBM and $\epsilon_F$ for electron doping at lower energy, than the AFM phase. In terms of one-electron-energy contribution, this means that either type of doping will lower the energy of the FM relative to the AFM phase: for hole doping, by removing valence electrons with higher energy from the FM than from the AFM bilayer, and for electron doping, by adding lower-energy conduction electrons to the FM than to the AFM bilayer. This is in agreement with the trends obtained for $E_\mathrm{ex}$ in the scalar-relativistic case, in figure~\ref{fig:Eex}, and consistent with the general expectation that weak doping favors ferromagnetism  \cite{Aka98}. 

In the fully-relativistic case, as far as the conduction band is concerned, the effect of changing from AFM to FM (figure~\ref{fig:DOS}(a) and (b)) is very similar to the scalar-relativistic case -- producing a shift to lower energy of the CBM (and associated electron-doping $\epsilon_F$ levels) nearly identical to that of the scalar relativistic situation. For the valence band, on the other hand, the behavior differs completely from the scalar-relativistic case, i.e., the VBM and $\epsilon_F$ for hole doping remain virtually unvaried upon changing from AFM to FM. In terms of band-energy contributions, this implies no variation with doping of the FM relative to the AFM energy -- which is fully consistent with the flat trend of $E_\mathrm{ex}$ in figure~\ref{fig:Eex}, but unlike the behavior generally expected for weak doping. In all cases thus the trends of $E_\mathrm{ex}$ with doping can be accounted for by the difference between the FM and AFM band-edge energy $\epsilon_F$ in figure~\ref{fig:DOS}.

\begin{figure*}[t]
\includegraphics[width=0.87\textwidth] {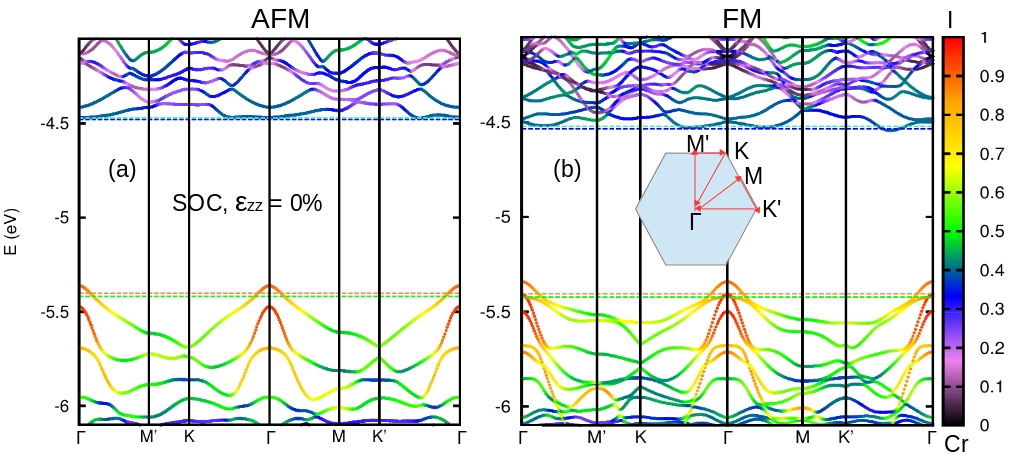}
\centering
	\caption{\label{fig:bands_pristine} Fully relativistic band structures of the pristine AFM (a) and FM (b) CrI$_{3}$ bilayer. The colors indicate the proportion of I and Cr  character in the projected band structures.
		The horizontal dashed lines show the Fermi level in the rigid-band model. The orange, green, blue and cyan dashed lines are for 0.1~h/cell, 0.2~h/cell, 0.1~e/cell and 0.2~e/cell doping, respectively. The energy scale is relative to the vacuum potential.}
\end{figure*}

The reason for the qualitatively different trend of the valence-band-edge energy $\epsilon_F$ in the relativistic case can be inferred from figure~\ref{fig:bands_pristine}, where we display the fully-relativistic atomic-projected band structures of the AFM and FM bilayers. In these band-structure plots, we indicate along each band the proportion of projected I and Cr atomic-orbital character -- we also reported, for easy reference, the corresponding $\epsilon_F$ energies of figure~\ref{fig:DOS}. The scalar-relativistic AFM and FM band structures are given in the Appendix, for comparison. Unlike the latter band structures (exhibiting {\bf k}-degenerate valence-band maxima outside $\Gamma$ and a quite-flat valence-band-edge region all around $\Gamma$), the fully-relativistic-band structure, in  figure~\ref{fig:bands_pristine}, shows a prominent single-{\bf k}-point VBM at the BZ center, with sharp band-edge curvature, both in the AFM and FM configurations. 

In addition, the atomic projections for the relativistic case, in figure~\ref{fig:bands_pristine}, evidence a quasi-total I-$5p$ atomic character of the upper four valence states at the BZ center (related to two 2-fold-degenerate bands in the AFM phase), consistent with relativistic results for the monolayer \cite{Lado_2017}. These states correspond to atomic I-$5p$ spin-orbital $|J=3/2, J_z = \pm 3/2 \rangle$ states, that have been pushed up into the bandgap (compared to the scalar-relativistic case, see Appendix) due to the very large I-$5p$ atomic spin-orbit splitting ($ \sim  0.9$~eV \cite{Min62,PetNie98}). The upper (lower) two of these valence bands in the central part of the BZ correspond to I-$5p$ states with moment  $J_z = + 3/2$ ($- 3/2$) parallel (antiparallel) to the magnetic moment of their CrI$_3$ layer. 

Hence, SOC completely modifies the shape of the valence-band edge and the location of the VBM positions in the BZ -- from a halo-type region around $\Gamma$ in the scalar-relativistic case to the $\Gamma$ point itself (figure~\ref{fig:bands_pristine}). 
Most importantly, the nature of the upper valence states and of the band broadening resulting in the VBM is different in the two cases. 
In the scalar-relativistic case (Appendix), the group of upper-four valence bands (two 2-fold-degenerate bands in the AFM phase) corresponds predominantly to Cr-$3d$ $t_{2g}$ out-of-plane states with considerable  I-$5p_z$ hybridization \cite{Wang_2016}. In that case, the substantial AFM to FM splitting/broadening of the corresponding group of bands outside $\Gamma$, due to interlayer interaction, effectively raises  the FM valence-edge energy $\epsilon_F$, relative to the AFM one (see Appendix). 
In the relativistic case, instead, the VBM at $\Gamma$ (of each CrI$_3$ monolayer) is a consequence of the strong I-$5p$ spin-orbit splitting. The corresponding relativistic I-$5p$ $|J=3/2, J_z = 3/2 \rangle$ VBM states are characterized by fully in-plane $p_x + i  p_y$ orbitals, which are unsuited to promote interaction between the two layers of the bilayer. This results in virtually identical AFM and FM  valence-edge energies $\epsilon_F$ in figure~\ref{fig:bands_pristine}. One can note, however, that in the rest of the BZ, where out-of-plane hybridization is promoted (with the Cr in particular), the FM broadening of the valence spectrum considerably increases.  

Based on the above results and rational, it can be expected that changing the position of the VBM away from $\Gamma$, where it is controlled by atomic-like relativistic states not sensitive to interlayer interactions, will enhance the valence-band broadening of the FM phase and therefore its stability (relative to the AFM phase) upon hole doping. In fact, we will show that this can be achieved effectively by applying moderate load/compression along the soft (out-of-plane) direction of the bilayer. Previously, it had been observed, in scalar-relativistic DFT+U calculations of the effect of in-plane biaxial strain on the bilayer band structure, that under 10\% stretching of the bilayer-surface area  (inducing a $d$ contraction), a displacement of the VBM from $\Gamma$ to outside the BZ center  was occurring \cite{LeoGonMej20}.

\subsection{\textbf{Overcoming the asymmetry of the electron and hole doping by vertical strain}}

\begin{figure*}[t]
\includegraphics[width=0.87\textwidth] {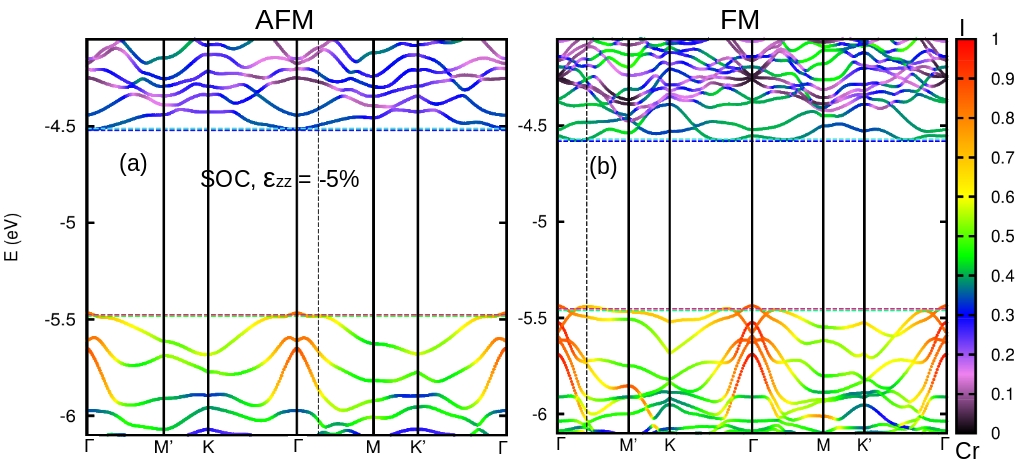}
\centering
	\caption{\label{fig:bands_strained} Fully relativistic bandstructures of  the   CrI$_{3}$ bilayer compressed in the perpendicular direction by 5\% for the AFM (a) and FM (b) configurations. The dashed vertical lines correspond to the new local VBM with highest energy appearing outside $\Gamma$ due to the application of the strain. The meaning of colors, horizontal lines and energy scale is described in the caption of figure~\ref{fig:bands_pristine}.  }
\end{figure*}

In figure~\ref{fig:bands_strained} we show the fully-relativistic atomic-projected band structures of the 5\%-compressed CrI$_3$ bilayer for the AFM and FM configurations. We note the AFM configuration remains lower in energy than the FM one under the 5\%-vertical compression. The compressive strain, in figure~\ref{fig:bands_strained}, is seen to lower the VBM energy at the BZ center and to induce new local valence-band maxima outside $\Gamma$, at some distance from the BZ center (see vertical dashed line in figure~\ref{fig:bands_strained}). 

The new (local) valence-band maxima induced by the strain are clearly much more pronounced in the FM configuration, particularly in the $\Gamma-M'$ direction, where the new VBM is slightly higher in energy than at $\Gamma$. The corresponding upper  valence band, outside the $\Gamma$ region, in the FM configuration is significantly lifted up in energy relative to the AFM one. In addition, the upper edge of this band, outside the BZ center, in the FM case is quite-flat over an extended halo-like region all around $\Gamma$; hence the DOS associated with the corresponding new valence band maxima clearly dominates compared to the one at $\Gamma$ (see also the corresponding DOS plots in the Supplementary Information). In fact, one observes, in figure~\ref{fig:bands_strained}, that contrary to the unstrained case, the Fermi levels for 0.1 and 0.2 hole/cell doping, in the rigid-band model, are shifted to higher energy in the FM compared to the AFM configuration. 
For the conduction band, on the other hand, the AFM-to-FM shifts in $\epsilon_F$  remain similar to the unstrained case. 
Therefore, for the strained bilayer, on the basis of the results in figure~\ref{fig:bands_strained}, one can expect the presence of an AFM-FM transition both with hole doping and with electron doping. 

\begin{figure*}[t]
\begin{center}
\includegraphics[width=0.47\textwidth] {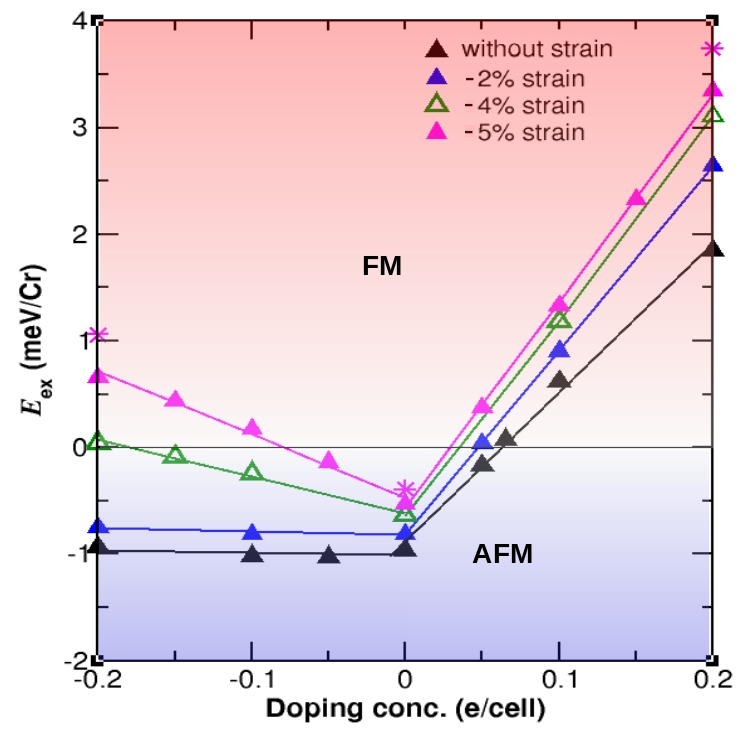}
\caption{\label{fig:Eex_strain} Exchange energy, $E_\mathrm{ex}$, as a function of doping for different  compressive uniaxial strains (2, 4 and 5\%) applied to the bilayer, compared to the unstrained bilayer. The stars denote the $E_\mathrm{ex}$ obtained by relaxing the bilayer in the presence of the dopant charge,  and this for the AFM and FM configurations separately, in the case of 5\% strain. All calculations were fully-relativistic. The same convention for hole and electron doping as in figure~\ref{fig:Eex} is used. }
\end{center}
\end{figure*}

Indeed, this is confirmed, in figure~\ref{fig:Eex_strain}, where we display the exchange energy $E_\mathrm{ex}$ of the strained bilayer, as a function of doping, for different values of the strain $\epsilon_{zz}$. 
In the bilayer, at sufficient vertical compression, an AFM to FM transition is induced both upon hole doping and upon electron doping.  
This occurs for $\epsilon_{zz} = -5$\% and $-4$\%. 
For the 5\% compression, the hole doping induces a transition at $\sim$0.08 hole/cell, while the transition upon electron doping occurs at 0.03~electron/cell, {\it i.e.,} a smaller doping value than in the unstrained bilayer.  
For 4\% compression, a higher level of hole doping (0.18 hole/cell) is necessary to induce the transition. 
For $\epsilon_{zz} = -2$\%, hole doping does not induce the transition.  At that strain, the valence-edge behavior is still controlled by the VBM at the BZ center, resulting in a constant $E_\mathrm{ex}$ with hole doping, as for the unstrained bilayer. 
One can notice, however, in figure~\ref{fig:Eex_strain}, that even in the absence of doping, compression smoothly increases $E_\mathrm{ex}$, and hence already in the 2\%-compressed bilayer the exchange coupling is somewhat less AFM than in the unstrained system. 

The AFM-FM transition with hole doping is enabled by the presence of the VBM feature away from $\Gamma$. 
It is in fact only under sufficient vertical strain, i.e., when the local VBM appears along $\Gamma - M'$ in the FM band structure (with the associated high DOS and rising energy relative to $\Gamma$), that the corresponding new local band edge starts enhancing the stability upon hole doping of the FM phase, compared to the AFM one. 
The corresponding band structures for 2\% and 4\% uniaxial strain are shown in  the Supplementary Information.  The slope, in general, of $E_\mathrm{ex}$ with hole (electron) doping, in figures~\ref{fig:Eex} and ~\ref{fig:Eex_strain}, is controlled by the energy difference between the band-edge $\epsilon_F$ positions in the AFM and FM configurations, $\Delta \epsilon_{F} = \epsilon_{F}^{AFM} - \epsilon_{F}^{FM}$. With stronger vertical compression, $\Delta \epsilon_F$ and hence the slope of $E_\mathrm{ex}$ vs. electron (hole) doping tend to increase in magnitude due to the increasing broadening outside $\Gamma$ of the FM conduction- (valence-) band edge relative to the AFM one. We note, in fact, that the band-energy offset $\Delta \epsilon_{F}$ per electron/hole, obtained from the AFM and FM rigid-band model with proper common electrostatic-potential reference level, accounts not only qualitatively, but surprisingly  even semi-quantitatively (within $\sim 20$~\% of the actual value) for the variation with doping of the exchange-coupling energy $E_\mathrm{ex}$ in all cases we considered here. 

Finally, we would like to comment on the origin of the new VBM feature in the strained bilayer. The new VBM feature derives from the rise in energy, upon vertical compression, of the two lower valence bands with main Iodine character in the central part of the BZ in figure~\ref{fig:bands_pristine} (at about --5.7 eV for both AFM and FM, see also DOS's in supplementary information). These two Iodine bands (which are degenerate in the AFM case) correspond to I-$5p$ J$=3/2$ states with moment $J_z = \pm 1/2$. Such states are moving up in energy,  upon $z$ compression, relative to the other Iodine bands with $|J_z| = 3/2$ in figure~\ref{fig:bands_pristine}, because of their major out-of-plane I-$p_z$ components (absent in the $|J_z| = 3/2$ states) -- becoming increasingly unfavorable energetically when the two inner planes of I anions in the bilayer are brought closer to each other. 
When the Iodine $|J_z| = 1/2$ bands of the central part of the BZ are shifted up in energy upon compression, states with different $J_z$ cannot interact essentially at the $\Gamma$ point and the $|J_z| = 1/2$ states can cross upper Iodine bands. However, when going away from $\Gamma$, they interact increasingly strongly, and the resulting level repulsion gives rise to the new VBM in figure~\ref{fig:bands_strained}.

\begin{figure*}[t]
\begin{center}
\includegraphics[width=0.55\textwidth]{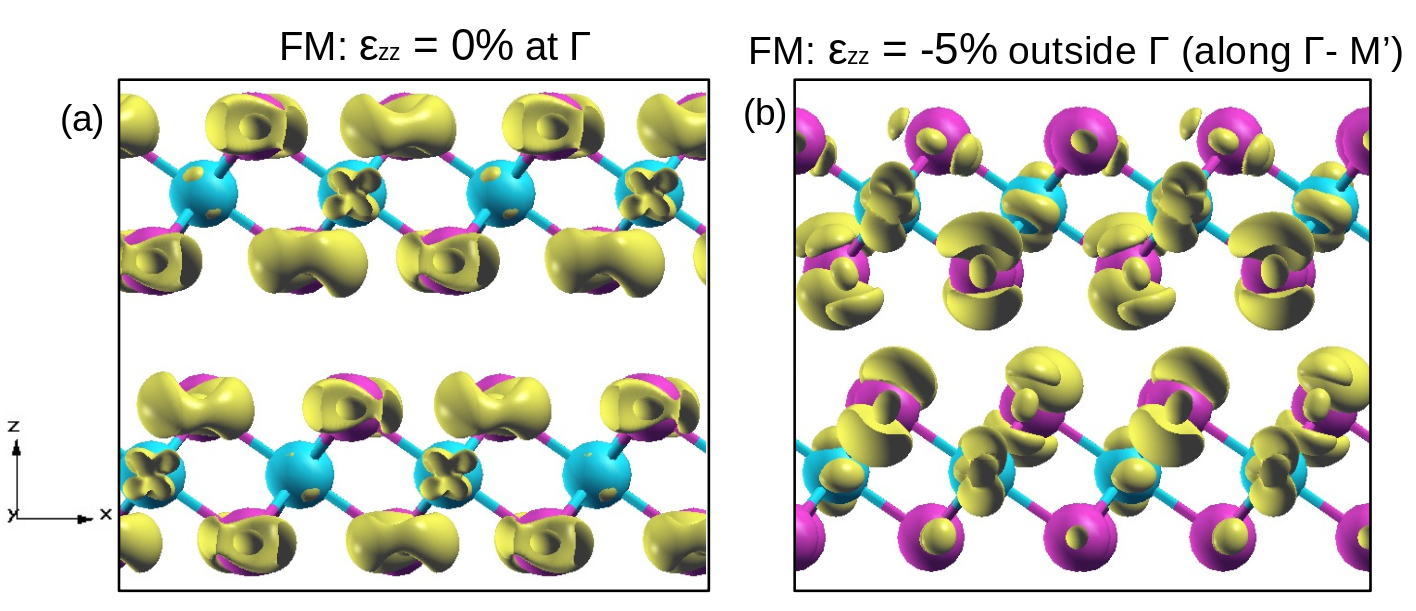}
\caption{\label{fig:states} Isosurface plots of the probability density of the VBM states for the pristine (a) and 5\%-compressed (b) FM bilayers. The VBM state is at $\Gamma$ for the pristine bilayer and outside $\Gamma$, in the $\mathrm \Gamma M'$ direction, for the strained bilayer.  Iso-surface value is $0.001$ e/Bohr$^{3}$.}
\end{center}
\end{figure*}

 In figure~\ref{fig:states}
we show the probability density of the states forming the VBM in the FM configuration for the pristine bilayer (VBM at $\Gamma$) (a) and  5\%-strained  bilayer (one of the two valence band maxima in the $\mathrm \Gamma M'$ direction) (b). 
In the pristine bilayer, as mentioned previously, the VBM corresponds to $|J=3/2, J_z = 3/2\rangle$  states of the Iodine atoms with in-plane $p_{\parallel}$ orbitals. The shape of the probability density in  figure~\ref{fig:states}(a) confirms that the orbitals are localized within the plane of the I atoms. Such relativistic valence-edge  states are responsible for the unusual trend with hole doping,  inhibiting the magnetic transition. 
On the contrary, in the strained bilayer (figure~\ref{fig:states}(b)), the VBM outside $\Gamma$, in the $\mathrm \Gamma  M'$ direction, corresponds predominantly to Iodine $|J=3/2, J_z = 1/2\rangle$ states with significant  $p_z$-orbital components; these orbitals extend in the space between the layers, favouring   interaction between the layers, and restoring the expected behavior for  $E_\mathrm{ex}$ that weak doping promotes ferromagnetism.

\section{Conclusions}

In this paper, using ab-initio density-functional-theory calculations, we found the origin of the asymmetric behavior of the hole doping compared to electronic doping
to lie in the strongly spin-orbit-split states forming the VBM at $\Gamma$. The specific relativistic nature of these atomic-like states at $\Gamma$ prevents the expected larger interactions and broadening for the FM bilayer configuration, which results in the suppression of the AFM-FM transition for hole doping. 


Furthermore, we showed that band engineering by application of moderate uniaxial strain perpendicular to the bilayer is able to lift the doping asymmetry and realize an AFM-FM transition also for hole doping. This is made possible thanks to the compressive-strain-induced creation of an additional VBM feature near $\Gamma$ which displays the expected broadening for the FM configuration and lowers its energy upon hole doping relative to the AFM configuration. We also explained the origin of this feature and the trend of the exchange coupling with combined doping and vertical strain.

Our results indicate that the combination of perpendicular strain and doping opens the possibility of controlling the magnetism by hole doping and also conveniently lowering the spin-flip field for potential switching applications in CrI$_{3}$-based heterostructures and related systems. More generally this study clarifies the factors which may be used to induce or inhibit the presence of a magnetic transition upon doping in the 2D layered-AFM insulators.

\section{Appendices}
\subsection{\textbf{Scalar-relativistic band-structure plots of CrI$_{3}$ bilayer}}

\begin{figure*}[t]
\begin{center}
\includegraphics[width=0.88\textwidth]
{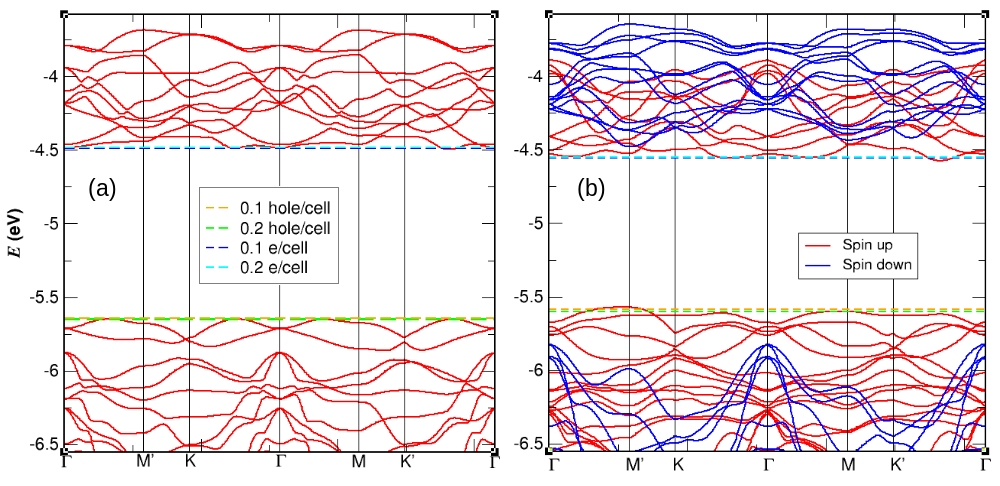}
\caption{\label{fig:Bands_AFM_FM} 
 Scalar-relativistic band structure of CrI$_{3}$ bilayer in (a) AFM and (b) FM configuration. The band energies are measured relative to the vacuum potential. The horizontal dashed lines show the Fermi level in the rigid-band model; the orange, green, blue and cyan dashed lines are for 0.1~h/cell, 0.2~h/cell, 0.1~e/cell and 0.2~e/cell doping, respectively. }
 \end{center}
\end{figure*}

The scalar-relativistic band-structure plots for the CrI$_{3}$ bilayer in AFM and FM configurations are shown in figure~\ref{fig:Bands_AFM_FM}. The band energies are given relative to the vacuum potential.  In both AFM and FM configurations, the valence band maxima appear outside $\Gamma$, which is a major  difference with respect to the fully-relativistic case.  For the AFM configuration, the VBM appears along the $\Gamma K(K'$) direction (see figure~\ref{fig:Bands_AFM_FM}(a)),  while for the FM configuration the VBM appears along the $\Gamma M'(M)$ direction     (see figure~\ref{fig:Bands_AFM_FM}(b)). The conduction band minima (CBM) appears along the $ \Gamma K(K'$) direction for both AFM and FM configurations.

From figure~\ref{fig:Bands_AFM_FM} we see that the valence band is significantly wider for the FM configuration than for the AFM one, due to the additional and rather large level splittings (outside $\Gamma$) induced by the interlayer interaction in the FM configuration. Therefore, as discussed in the main text, this favors the FM arrangement upon hole doping. 

\subsection{\textbf{ Energy barrier for the  monoclinic to rhombohedral transition in bilayer CrI$_{3}$}}

In order to confirm that the application of perpendicular compression does not alter the (meta-)stable character of the bilayer monoclinic phase, 
 we evaluated the transition barrier between the HT-AFM phase and the LT-FM phases, with and without strain. For the unstrained bilayer, previous DFT calculations \cite{JiaWanChe19} have shown an energy barrier of about 10 meV/Cr for the structural transformation from the monoclinic high-temperature (HT) to rhombohedral low-temperature (LT) phase.
 
\begin{figure*}[b]
\begin{center}
\includegraphics[width=0.58\textwidth]
{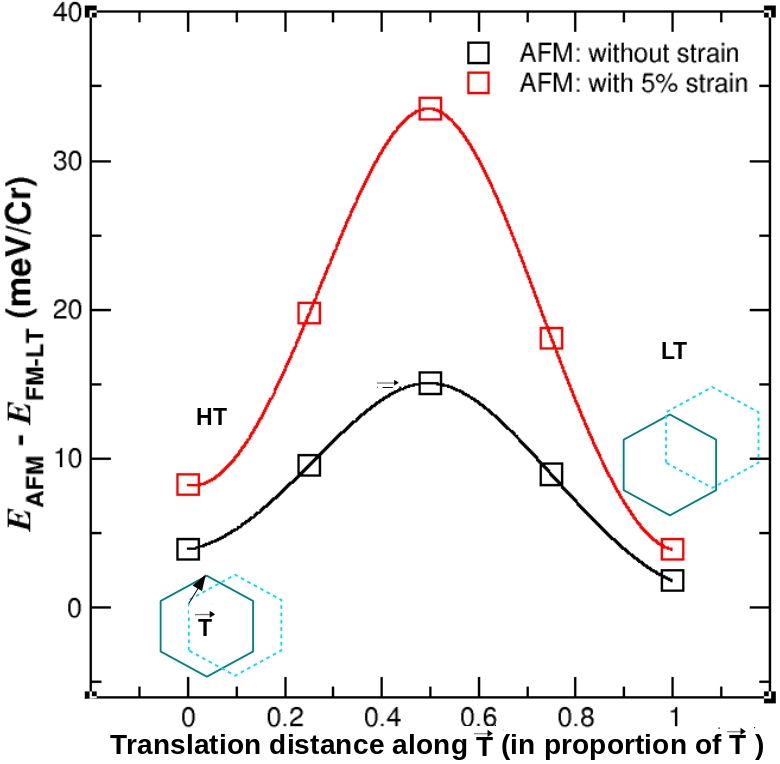}
\caption{\label{fig:stacking}Energy barrier for the transition from the high-temperature (HT) AFM to the low-temperature (LT) FM phase of the unstrained and 5\%-compressed bilayer. One of the monolayers of the AFM bilayer is translated along the $\rm \vec{T}$ vector (see inset) relative to the other layer. The translation is expressed as proportion of the distance between the HT and LT sites and is given along the horizontal axis. Along the vertical axis we plot the energy of the AFM bilayer, for different stacking configurations (obtained by sliding the monolayer from HT to LT site), relative to the energy of the LT FM bilayer.
}
\end{center}
\end{figure*}

 To reach the LT phase from the HT phase, we have  gradually translated one of the layers  relative to the other layer along the vector $\vec{T}$, joining the HT and LT site (see inset in figure~\ref{fig:stacking}). 
The energy of the corresponding bilayer in the AFM arrangement, measured relative to the energy of the LT-FM bilayer, is given in figure~\ref{fig:stacking} as a function of the sliding, both for the pristine and strained bilayer. For the sliding, we fixed the positions of Cr atoms and relaxed the positions of all I atoms. 
  For the pristine bilayer, the energy difference between the HT-AFM and LT-FM phase is 3.9~meV/Cr and the energy barrier from the HT-AFM to the LT-FM structure amounts to 11.8~meV/Cr, in good agreement with previous DFT-GGA results \cite{JiaWanChe19}.   In the presence of  5\%-compressive strain along the out-of-plane direction the energy difference increases to 8.2~ meV/Cr and the energy barrier  to  25.2~meV/Cr. The transition barrier remains virtually unchanged if  one gradually translates the HT-FM layers, instead of the HT-AFM layers. 

Hence, compared to the unstrained system, the large increase in the energy barrier we find here with the vertical compressive strain will tend to further inhibit a spontaneous structural transition from the HT to LT phase. The experimental bilayer is expected thus also to remain in the HT phase with monoclinic stacking under the type of compressive strains we consider in this work.

\subsection{\label{sec:Append_strain}\textbf{ Uniaxial versus biaxial strain on CrI$_{3}$ bilayer}}

We have performed {\it ab initio} calculations to compare the overall bilayer deformations (aspect ratio) induced by uniaxial vertical and biaxial in-plane strains. For the  uniaxial strain, we compressed the bilayer by 5\% in the perpendicular direction, and obtained a 1\% expansion of the in-plane lattice constant: ($\epsilon_{zz}/\epsilon_{xx})_{\bot}= -5$, where $\epsilon_{ij}$ stands for the $ij$ component of the strain tensor. For the biaxial in-plane strain, we expanded the in-plane lattice parameter by 1\%, and obtained a 0.5\% contraction in the perpendicular direction:   ($\epsilon_{zz}/\epsilon_{xx})_{\parallel} = -0.5$. Thus, for the same expansion of the in-plane lattice parameter, the bilayer thickness varies by a factor of 10, depending on the type of strain imposed. We checked, using different strain values, that these results correspond to the linear elastic regime. 

The above large difference (factor of 10) in the deformation ratio is related to the van der Waals nature of the CrI$_3$ system and resulting elastic modulii. In the case of uniaxial strain applied in the perpendicular direction of a crystal with hexagonal Bravais lattice, within linear elasticity, the strain ratio can be written as:  
\begin{equation}
     \left(\frac{\epsilon_{zz}}{\epsilon_{xx}}\right )_\bot = -\frac{c_{11}+c_{12}}{c_{13}},
     \label{eq2}
\end{equation}
where $c_{ij}$ are components of the crystal elastic tensor. 
For biaxial in-plane strain, the strain ratio becomes:
\begin{equation}
     \left(\frac{\epsilon_{zz}}{\epsilon_{xx}}\right)_{\parallel}  = -\frac{2c_{13}}{c_{33}}.
     \label{eq4}
\end{equation}
Typically, in van der Waals systems, like graphite, the ratio $c_{11}/c_{13}$ tends to be one or more orders of magnitude larger than the ratio $c_{13}/c_{33}$ \cite{BlaProSel70}. This leads to a considerably larger strain ratio in Eq.~(\ref{eq2}) than in Eq.~(\ref{eq4}), which, e.g., in graphite, can be as much as 100 times larger \cite{BlaProSel70}:  $ \left(\frac{\epsilon_{zz}}{\epsilon_{xx}}\right )_\bot \approx 100  \left(\frac{\epsilon_{zz}}{\epsilon_{xx}}\right)_{\parallel} $.

Under the 5\% uniaxial perpendicular compression of the CrI$_3$ bilayer, the separation between the two inner I atomic planes of the bilayer is reduced from $d_{\rm int}^{\rm eq}=3.36$~\AA\  to $d_{\rm int}=3.07$~\AA. In the absence of strain, within each CrI$_3$ layer, the separations between the bilayer outer I plane and Cr plane ($d_1$) and between the inner I plane and Cr plane ($d_2$) are nearly identical (differ by less than 0.1 \%; $d_1 = d_2 = 1.53$~\AA). Under the 5\% compression along $z$, $d_1$ decreases by 3.2\% and $d_2$ by 2.4\%. These changes and their difference are much stronger than the ones resulting from the application of large  electric fields perpendicular to the monolayer \cite{GhoStoBin19}. 




\section{References}

\bibliography{iopart-num}

\providecommand{\newblock}{}
\begin{thebibliography}{10}
\expandafter\ifx\csname url\endcsname\relax
  \def\url#1{{\tt #1}}\fi
\expandafter\ifx\csname urlprefix\endcsname\relax\def\urlprefix{URL }\fi
\providecommand{\eprint}[2][]{\url{#2}}

\bibitem{SONG201733}
Song C, Cui B, Li F, Zhou X and Pan F 2017 {\em Prog. Mater. Sci.\/} {\bf 87}
  33--82

\bibitem{Matsukura2015}
Matsukura F, Tokura Y and Ohno H 2015 {\em Nat. Nanotechnol.\/} {\bf 10}(3)
  209--220

\bibitem{GonLiLi17}
Gong C, Li L, Li Z, Ji H, Stern A, Xia Y, Cao T, Bao W, Wang C, Wang Y, Qiu
  Z~Q, Cava R~J, Louie S~G, Xia J and Zhang X 2017 {\em Nature\/} {\bf 546} 265

\bibitem{HuangNature2017}
Huang B, Clark G, Navarro-Moratalla E, Klein D~R, Cheng R, Seyler K~L, Zhong D,
  Schmidgall E, McGuire M~A, Cobden D~H, Yao W, Xiao D, Jarillo-Herrero P and
  Xu X 2017 {\em Nature\/} {\bf 546} 270--273

\bibitem{GibKop19}
Gibertini M, Koperski M, Morpurgo A~F and Novoselov K~S 2019 {\em Nat.
  Nanotechnol.\/} {\bf 14} 408

\bibitem{Tsymbal2019}
Paudel T~R and Tsymbal E~Y 2019 {\em ACS Appl. Mater. Interfaces\/} {\bf 11}
  15781--15787

\bibitem{Nardelli2020}
Heath J~J, Costa M, Buongiorno-Nardelli M and Kuroda M~A 2020 {\em Phys. Rev.
  B\/} {\bf 101}(19) 195439

\bibitem{Stroppa2020}
Hu T, Zhao G, Gao H, Wu Y, Hong J, Stroppa A and Ren W 2020 {\em Phys. Rev.
  B\/} {\bf 101} 125401

\bibitem{DolPetZol20}
Dolui K, Petrović M~D, Zollner K, Plecháč P, Fabian J and Nikolić B~K 2020
  {\em Nano Lett.\/} {\bf 20} 2288

\bibitem{GonZha19}
Gong C and Zhang X 2019 {\em Science\/} {\bf 363} 4450

\bibitem{Nanoscale}
Zheng F, Zhao J, Liu Z, Li M, Zhou M, Zhang S and Zhang P 2018 {\em
  Nanoscale\/} {\bf 10}(29) 14298--14303

\bibitem{McGuire2015}
McGuire M~A, Dixit H, Cooper V~R and Sales B~C 2015 {\em Chem. Mater.\/} {\bf
  27} 612--620

\bibitem{Ubrig_2019}
Ubrig N, Wang Z, Teyssier J, Taniguchi T, Watanabe K, Giannini E, Morpurgo A~F
  and Gibertini M 2019 {\em 2D Mater.\/} {\bf 7} 015007

\bibitem{Thiel973}
Thiel L, Wang Z, Tschudin M~A, Rohner D, Guti{\'e}rrez-Lezama I, Ubrig N,
  Gibertini M, Giannini E, Morpurgo A~F and Maletinsky P 2019 {\em Science\/}
  {\bf 364} 973--976

\bibitem{JiaShaMak18}
Jiang S, Shan J,  and Mak K~F 2018 {\em Nat. Mater.\/} {\bf 17}(5) 406--410

\bibitem{HuaClaKle18}
Huang B, Clark G, Klein D~R, MacNeill D, Navarro-Moratalla E, Seyler K~L,
  Wilson N, McGuire M~A, Cobden D~H, Xiao D, Yao W, Jarillo-Herrero P and Xu X
  2018 {\em Nat. Nanotechnol.\/} {\bf 13}(7) 544--548

\bibitem{JiaXieSha20}
Jiang S, Xie H, Shan J and Mak K~F 2020 {\em Nat. Mater.\/} {\bf 19} 1295

\bibitem{Song2019}
Song T, Fei Z, Yankowitz M, Lin Z, Jiang Q, Hwangbo K, Zhang Q, Sun B,
  Taniguchi T, Watanabe K, McGuire M~A, Graf D, Cao T, Chu J~H, Cobden D~H,
  Dean C~R, Xiao D and Xu X 2019 {\em Nat. Mater.\/} {\bf 18}(12) 1298--1302

\bibitem{Li2019}
Li T, Jiang S, Sivadas N, Wang Z, Xu Y, Weber D, Goldberger J~E, Watanabe K,
  Taniguchi T, Fennie C~J, Fai~Mak K and Shan J 2019 {\em Nat. Mater.\/} {\bf
  18}(12) 1303--1308

\bibitem{LeoGonMej20}
León A~M, González J~W, Mejía-López J, de~Lima F~C and Morell E~S 2020 {\em
  2D Mater.\/} {\bf 7} 035008 and its Supplementary information

\bibitem{SivOkaXu18}
Sivadas N, Okamoto S, Xu X, Fennie C~J and Xiao D 2018 {\em Nano Lett.\/} {\bf
  18}(12) 7658--7664

\bibitem{JiaWanChe19}
Jiang P, Wang C, Chen D, Zhong Z, Yuan Z, Lu Z~Y and Ji W 2019 {\em Phys. Rev.
  B\/} {\bf 99}(14) 144401--9

\bibitem{SorCarFer19}
Soriano D, Cardoso C and Fernández-Rossier J 2019 {\em Solid State Commun.\/}
  {\bf 299} 113662

\bibitem{Su_rez_Morell_2019}
Morell E~S, Le{\'{o}}n A, Miwa R~H and Vargas P 2019 {\em 2D Mater.\/} {\bf 6}
  025020

\bibitem{XuZou20}
Xu R and Zou X 2020 {\em J. Phys. Chem. Lett.\/} {\bf 11}(8) 3152--3158

\bibitem{SorKat20}
Soriano D and Katsnelson M~I 2020 {\em Phys. Rev. B\/} {\bf 101}(4) 041402--5

\bibitem{LeiChi19}
Lei C, Chittari B~L, Nomura K, Banarjee N, Jung J and MacDonald H~A 2019
  Preprint at https://arxiv.org/abs/1902.06418

\bibitem{LiuZhoZha20}
Liu N, Zhou S and Zhao J 2020 {\em Phys. Rev. Mater.\/} {\bf 4} 094003

\bibitem{JiaLiWan18}
Jiang S, Li L, Wang Z, Mak K~F and Shan J 2018 {\em Nat. Nanotechnol.\/} {\bf
  13}(7) 549--553

\bibitem{Aka98}
Akai H 1998 {\em Phys. Rev. Lett.\/} {\bf 81} 3002

\bibitem{GiaBarBon09}
Giannozzi P, Baroni S, Bonini N, Calandra M, Car R, Cavazzoni C, Ceresoli D,
  Chiarotti G~L, Cococcioni M, Dabo I, Corso A~D, de~Gironcoli S, Fabris S,
  Fratesi G, Gebauer R, Gerstmann U, Gougoussis C, Kokalj A, Lazzeri M,
  Martin-Samos L, Marzari N, Mauri F, Mazzarello R, Paolini S, Pasquarello A,
  Paulatto L, Sbraccia C, Scandolo S, Sclauzero G, Seitsonen A~P, Smogunov A,
  Umari P and Wentzcovitch R~M 2009 {\em J. Condens. Matter Phys."\/} {\bf 21}
  395502

\bibitem{GhoStoBin19}
Ghosh S, Stoji{\' c} N and Binggeli N 2019 {\em Physica B: Cond. Matt.\/} {\bf
  570} 166

\bibitem{comment_SG}
Ghosh S, Stojić N and Binggeli N 2020 {\em Appl. Phys. Lett.\/} {\bf 116}
  086101

\bibitem{Lado_2017}
Lado J~L and Fern{\'{a}}ndez-Rossier J 2017 {\em 2D Mater.\/} {\bf 4} 035002

\bibitem{Min62}
Minnhagen L 1962 {\em Ark. Fys.\/} {\bf 21} 415

\bibitem{PetNie98}
Pettersson M and Nieminen J 1998 {\em Chem. Phys. Lett.\/} {\bf 283} 1 -- 6

\bibitem{Wang_2016}
Wang H, Fan F, Zhu S and Wu H 2016 {\em Europhys. Lett.\/} {\bf 114} 47001

\bibitem{BlaProSel70}
Blakslee O~L, Proctor D~G, Seldin E~J, Spence G~B and Weng T 1970 {\em J. Appl.
  Phys.\/} {\bf 41} 3373--3382

\end{thebibliography}

\newpage

\begin{center}
{\textbf{\large{Supporting Information: \\
Overcoming the asymmetry of the electron and hole doping for magnetic transitions in bilayer CrI$_{3}$
}}}

\end{center}


\setcounter{figure}{0}

\setcounter{section}{0}

\renewcommand{\thesection}{S\arabic{section}}  
\section{\textbf{Planar average plots for electrostatic potential of CrI$_{3}$ bilayer without and with 5\% uniaxial compressive strain}}

\begin{figure*}[h]
\includegraphics[width=0.95\textwidth]
{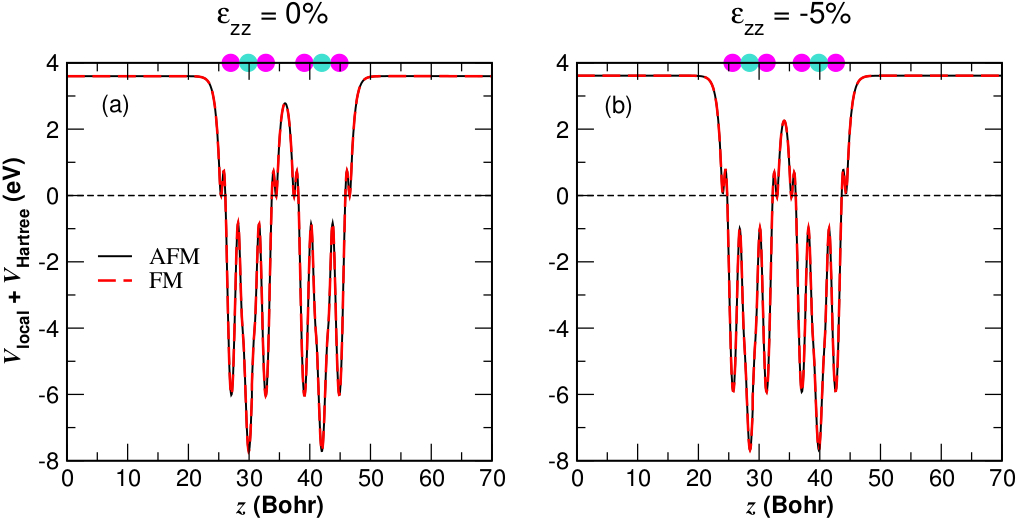}
\renewcommand{\thefigure}{S\arabic{figure}}  
\caption{\label{Potential} Planar average of the electrostatic ($V_\mathrm{local}+V_\mathrm{Hartree}$) potential in the supercell for the (a) unstrained ($\epsilon_\mathrm{zz}=0\%$) and (b) strained ($\epsilon_\mathrm{zz}=-5\%$) CrI$_{3}$ bilayers in AFM (black curve) and FM (red dashed curve) configurations. The magenta and turquoise spheres show the positions of the iodine and chromium atoms, respectively, along the $z$ direction in the supercell. }
\end{figure*}

The planar averages of $V_\mathrm{local}+V_\mathrm{Hartree}$ along the out-of-plane ($z$) direction for the unstrained and strained ($\epsilon_\mathrm{zz}=-5\%$) CrI$_{3}$ bilayers in AFM and FM configurations are shown by the black and red curves, respectively, in  figure~\ref{Potential}.  The energy position of the vacuum level is virtually identical for AFM and FM configurations. The difference in vacuum level between AFM and FM configurations is $< 0.05$ meV. The maximum difference of the electrostatic potential (in the vicinity of the atomic layers) between the AFM and FM configurations is $\approx$ 1 meV.
These are consequences of the total charge density (electronic + ionic) of the AFM and FM bilayers being nearly identical. 


\section{\textbf{Band structures of the  doped  CrI$_{3}$ bilayer at $\epsilon_\mathrm{zz}=0\%$ supporting rigid-band model}}

\begin{figure}[t]
\begin{center}
\includegraphics[width=0.95\textwidth]
{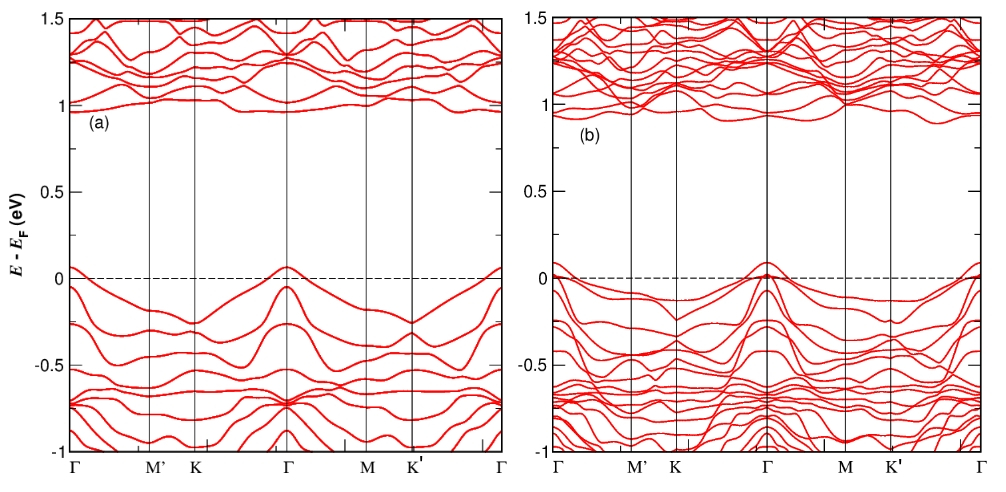}
\renewcommand{\thefigure}{S\arabic{figure}}
\caption{\label{BS_doped_unstrained}Fully-relativistic band structure plots of the doped CrI$_{3}$ bilayer in (a) AFM  and (b) FM configurations without any strain ($\epsilon_\mathrm{zz}=0\%$) at 0.1 hole/cell doping. The horizontal dashed line corresponds to the Fermi energy.  }
\end{center}
\end{figure}

Relativistic band structure of bilayer CrI$_{3}$ in AFM and FM  configurations obtained from self-consistent charge-density  calculations with 0.1~hole/cell doping is shown in figure~\ref{BS_doped_unstrained}. The band structure of the doped system and position of the Fermi level in the bands are similar to the results obtained with the rigid-band model of the pristine bilayer, both for the AFM and FM configurations. 

\section{\textbf{Band structure of CrI$_{3}$ bilayer with 2\% and 4\% uniaxial compressive strain }}

\begin{figure}[t]
\begin{center}
\includegraphics[width=0.95\textwidth]
{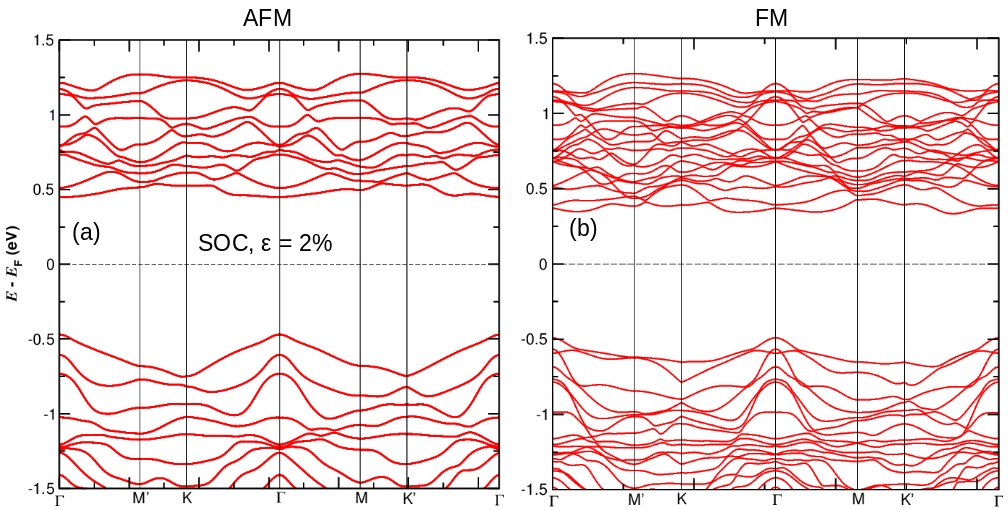}
\renewcommand{\thefigure}{S\arabic{figure}}
\caption{\label{Bands_SOC_2}Fully-relativistic band structure of CrI$_{3}$ bilayer under (a), (b) 2\%  uniaxial compressive strain in AFM and FM configurations.} 
\end{center}
\end{figure}

\begin{figure}[t]
\begin{center}
\includegraphics[width=0.95\textwidth]
{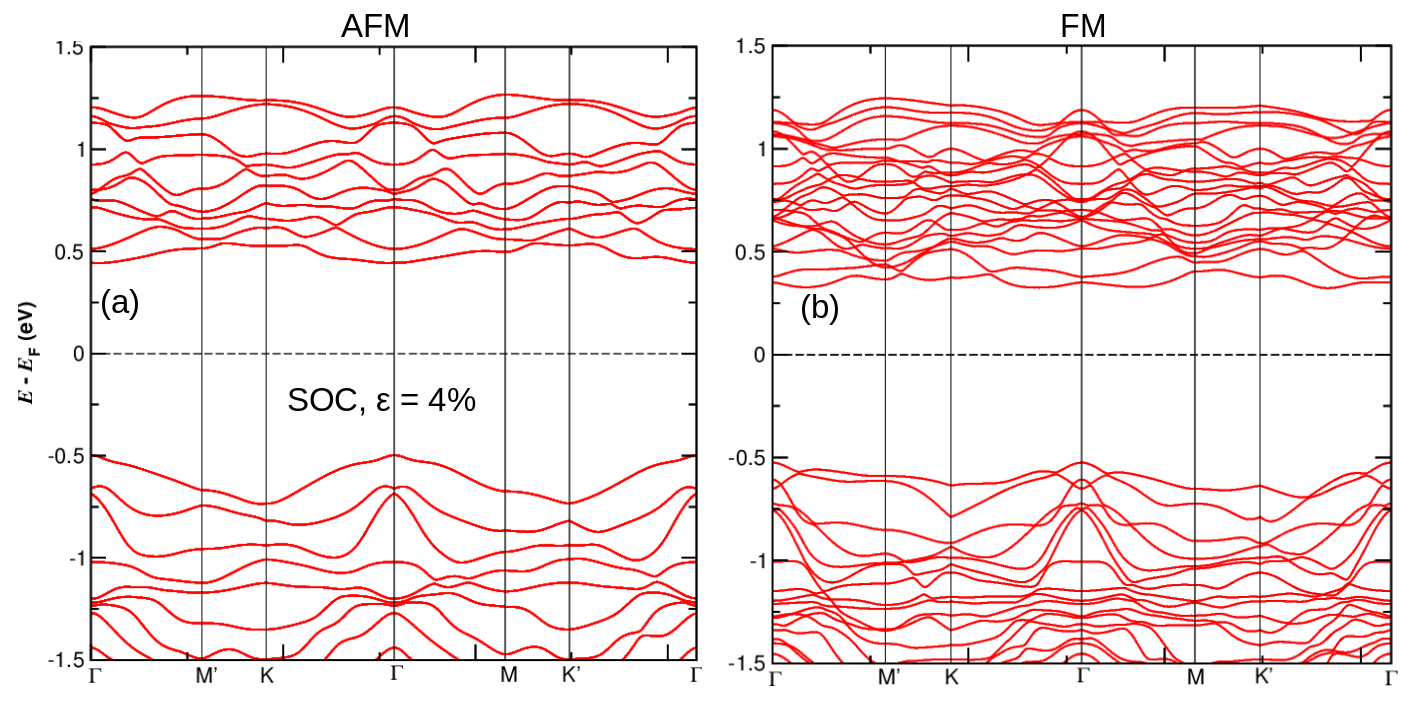}
\renewcommand{\thefigure}{S\arabic{figure}}
\caption{\label{Bands_SOC_4}Fully-relativistic band structure of CrI$_{3}$ bilayer under (a), (b) 4\%  uniaxial compressive strain in AFM and FM configurations.} 
\end{center}
\end{figure}

Fully-relativistic band structure plots of CrI$_{3}$ bilayer in AFM configuration with 2\% and 4\% compressive strain (applied along out-of-plane direction) are shown in Figs.~\ref{Bands_SOC_2} and ~\ref{Bands_SOC_4}. The reduction of the dispersion of the highest valence band in the central part of the BZ and beginning of creation of a new VBM  feature around $\Gamma$ can be observed  for increase of the strain from  2 to 4\%, in agreement with the trends observed in figures~4 and 5 in the main manuscript.

\section{\textbf{Electronic density of states for CrI$_{3}$ bilayer with 0\% and 5\% uniaxial compressive strain }}

\begin{figure}[t]
\begin{center}
\includegraphics[width=0.82\textwidth]
{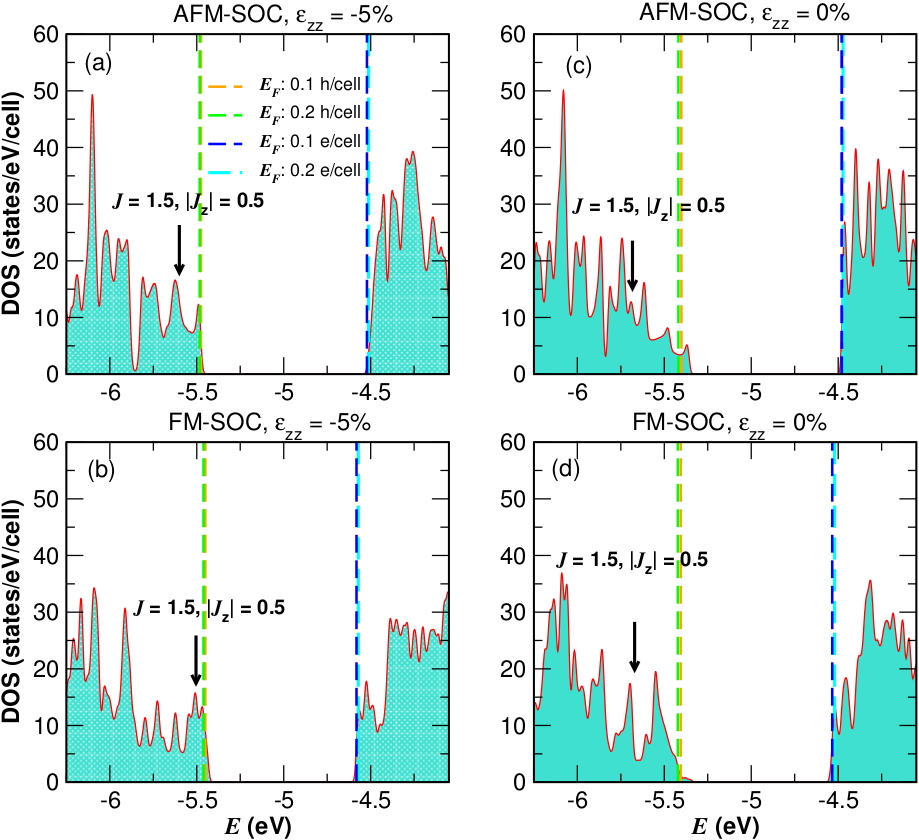}
\renewcommand{\thefigure}{S\arabic{figure}}
\caption{\label{DOS_5percent} Electronic density of states of bilayer CrI$_{3}$ (including SOC) under (a)--(b) 0\% and (c)--(d) 5\% compressive uniaxial strain for (a), (c) AFM and (b), (d) FM configurations. The energy is given with respect to the vacuum potential. The dashed orange, green, blue and cyan lines are the Fermi energies corresponding to 0.1 hole/cell, 0.2 hole/cell, 0.1 e/cell and 0.2 e/cell doping, respectively, obtained from rigid-band model. The vertical black arrow in each plot shows the energy level with $J = 1.5, |J_{z}| = 0.5$ (pointing by the black arrow).  The uniaxial compressive strain causes the $|J_{z}| = 0.5$ states in the valence band to rise up in energy.}
\end{center}
\end{figure}

Fig.~\ref{DOS_5percent} shows the fully-relativistic electronic density of states (DOS) for the unstrained and strained (5\% compressive strain) CrI$_{3}$ bilayer in AFM and FM configurations.
In comparison to DOS of the pristine bilayer in Fig.~\ref{DOS_5percent} (a) and (b), a large increase of DOS at the VBM can be seen upon the compression, see Figs.~\ref{DOS_5percent}(c) and (d), especially in the FM case. Furthermore, it can be observed that the broadening of the FM valence and conduction bands is larger than for the AFM configuration, supported by the positions of Fermi level for 0.1 e(h) doping.  Similar conclusion was reached for the band structures shown in figure~5. The uniaxial compressive strain causes the $J_{z}=|0.5|$ states to shift towards higher energy in the valence band, i.e., moves close to the Fermi energy, as indicated by the arrows in  Fig. ~\ref{DOS_5percent}.

\section{\textbf{Validation of rigid-band model for  doped CrI$_{3}$ bilayer at $\epsilon_\mathrm{zz}=-5\%$}}

\begin{figure}[t]
\includegraphics[width=0.95\textwidth]
{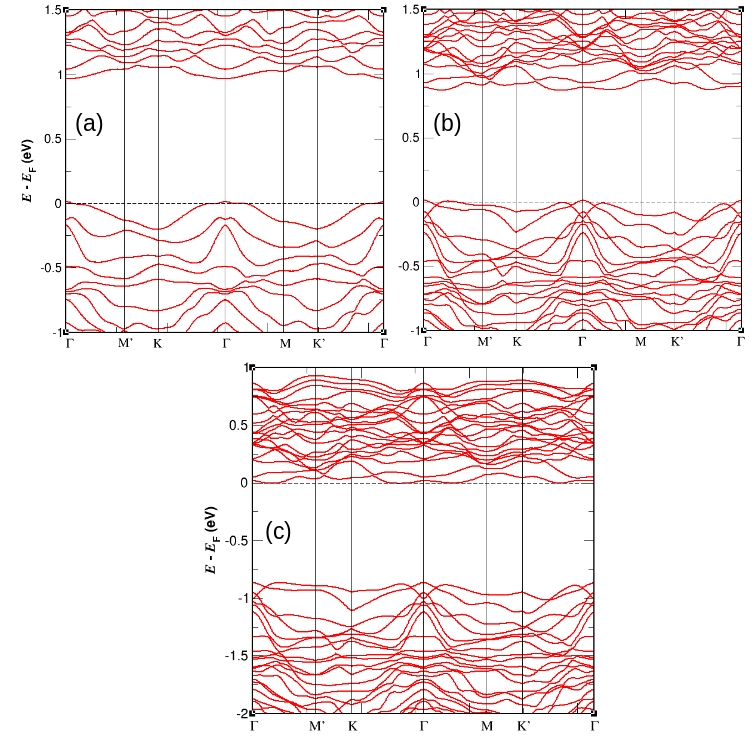}
\renewcommand*{\thefigure}{S\arabic{figure}}
\caption{\label{Bands_SOC_5_AFM_FM_hole} Fully-relativistic band structure of doped CrI$_{3}$ bilayer under 5\% compressive strain: (a) AFM  and (b) FM configurations with 0.1 hole/cell doping, and (c) FM configuration at 0.1 electron/cell doping.  The horizontal dashed line in each plot corresponds to the Fermi energy. }
\end{figure}

Fully-relativistic band structure of bilayer CrI$_{3}$ in AFM  and FM  configurations under 5\% compressive strain from self-consistent charge-density calculations with 0.1~hole/cell doping is shown in figures~\ref{Bands_SOC_5_AFM_FM_hole} (a) and (b). The corresponding band structure in FM configuration for 0.1 electron/cell doping at $\epsilon_\mathrm{zz}=-5\%$ is shown in figure~\ref{Bands_SOC_5_AFM_FM_hole}(c). 
The band structures of the doped systems and positioning of the Fermi level in the bands are similar to the results obtained with the undoped bilayer and rigid-band model, both for the AFM and FM configurations.

\section{\textbf{3D isosurface and planar average plots for the probability density of the upper valence states in the pristine and strained bilayer}}

\begin{figure*}[h]
\begin{center}
\includegraphics[width=0.95\textwidth]
{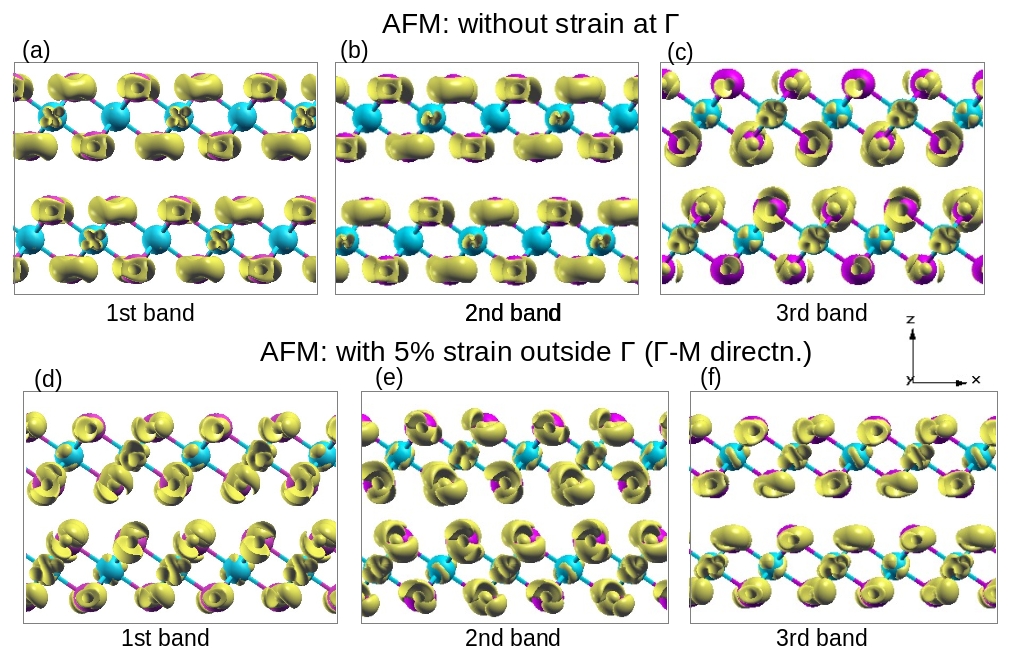}
\renewcommand*{\thefigure}{S\arabic{figure}}
\caption{\label{3D_AFM} 3D iso-surface plots of the probability density for the upper valence states of CrI$_{3}$ bilayer in AFM configuration. (a)--(c) Unstrained bilayer at $\Gamma$ and (d)--(f) strained (5\%) bilayer outside $\Gamma$ along $\Gamma-M$ direction. (a) and (d): first upper valence band; (b) and (e): second upper valence band; (c) and (f): third upper valence band for the unstrained and strained CrI$_{3}$ bilayers, respectively. Drawn at iso-surface value = 0.002 e/Bohr$^{3}$.}
\end{center}
\end{figure*}

Figure~\ref{3D_AFM} shows the 3D isosurface plots for the probability density, $|\Psi|^{2}$, of the upper three valence bands in AFM pristine configuration (a)--(c) and with 5\% compression (d)--(f).
In the case of the unstrained bilayer $|\Psi|^{2}$ for the upper three valence bands are plotted at $\Gamma$, corresponding to the VBM position.  Since the local VBM for the 5\% compression is  along $\Gamma-M$ direction, outside $\Gamma$, we plotted $|\Psi|^{2}$ for the upper valence bands at the position of the new local VBM. 
The highest 2 bands (figure~\ref{3D_AFM} (a-b)) 
in the pristine bilayer display predominant in-plane character of the states located on the I atoms, while the probability density for the third band (figure~\ref{3D_AFM} (c)) shows much more out-of-plane features.  Upon addition of the compressive perpendicular strain, the local VBM is characterized by states with more pronounced out-of-plane character on the atoms of the inner I planes of the bilayer (figure~\ref{3D_AFM} (a-b)).

\begin{figure}[h]
\begin{center}
\includegraphics[width=0.95\textwidth]{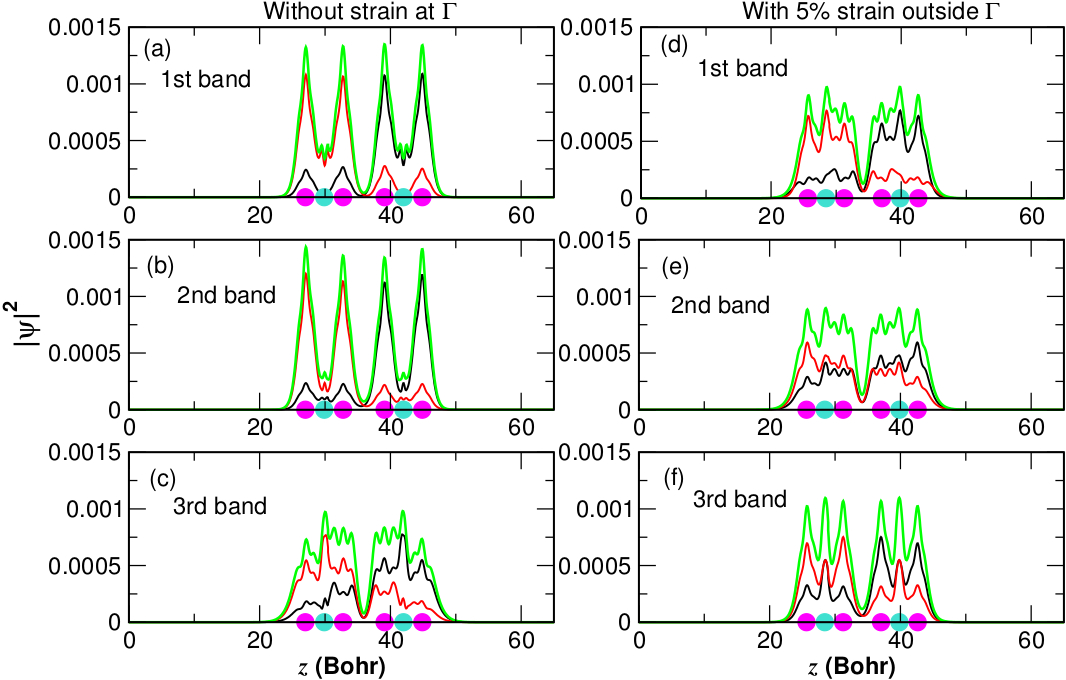}
\renewcommand{\thefigure}{S\arabic{figure}}
\caption{\label{AFM_plavg}Planar average plots of the probability density for the first three upper valence bands of the pristine and strained AFM CrI$_{3}$ bilayer. (a)--(c) Unstrained bilayer at $\Gamma$ and (d)--(f) compressed (5\%) bilayer outside $\Gamma$ (along $\Gamma-M$ direction). (a) and (d) First upper valence band; (b) and (e) second upper valence band; (c) and (f) third upper valence band for the unstained and strained systems, respectively. The black and red curves correspond to the degenerate states belonging to a particular band, the green curve is the sum. }
\end{center}
\end{figure}

Figure~\ref{AFM_plavg} shows the planar average plots for $|\Psi|^{2}$ of the upper three valence bands in the pristine and strained AFM configuration. It can be observed that in the pristine bilayer the highest lying two bands have the states localized on one of the layers  of the bilayer, on the I atoms. 
The third band, on the other hand shows a stronger probability density in the Cr region and much more extended probability density in the space between the layers.  The states of the highest two bands in the strained bilayer, outside $\Gamma$,  bear significant resemblance to the states of the third-upper band at $\Gamma$ in the unstrained bilayer. 

\begin{figure}[h]
\begin{center}
\includegraphics[width=0.95\textwidth]{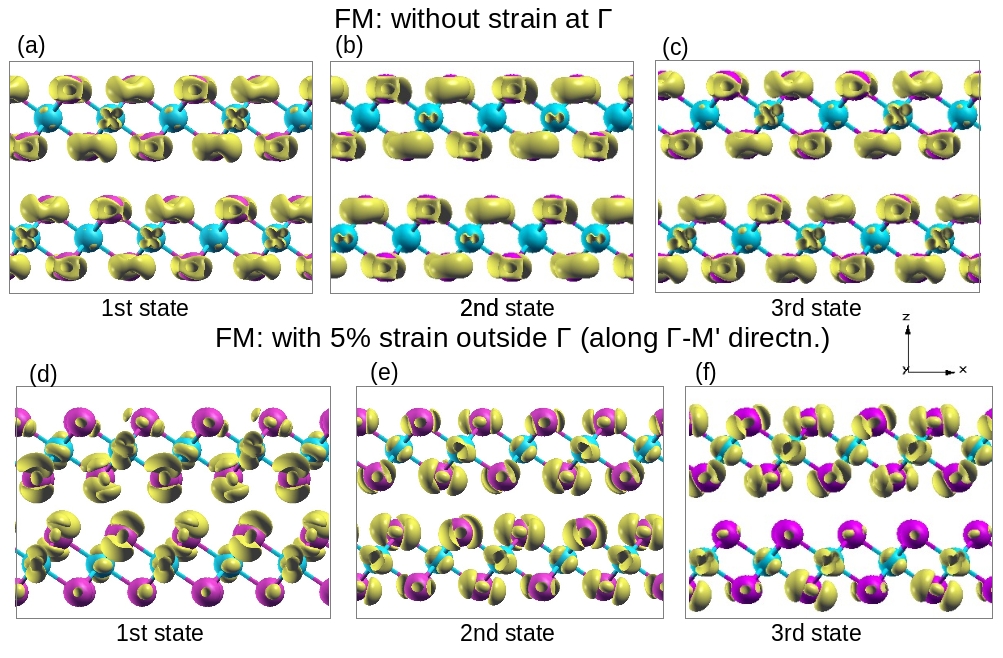}
\renewcommand{\thefigure}{S\arabic{figure}}
\caption{\label{3D_FM_first} 3D iso-surface plots of the probability density for the upper valence states of CrI$_{3}$ bilayer in FM configuration. (a)--(c) Unstrained bilayer at $\Gamma$, (d)--(f) strained (5\%) bilayer outside $\Gamma$ along $\Gamma-M'$ direction. (a) and (d): first upper valence state; (b) and (e): second upper valence state; (c) and (f): third upper valence state. Drawn at iso-surface value = 0.001 e/Bohr$^{3}$.}
\end{center}
\end{figure}

\begin{figure}[h]
\begin{center}
\includegraphics[width=0.95\textwidth]{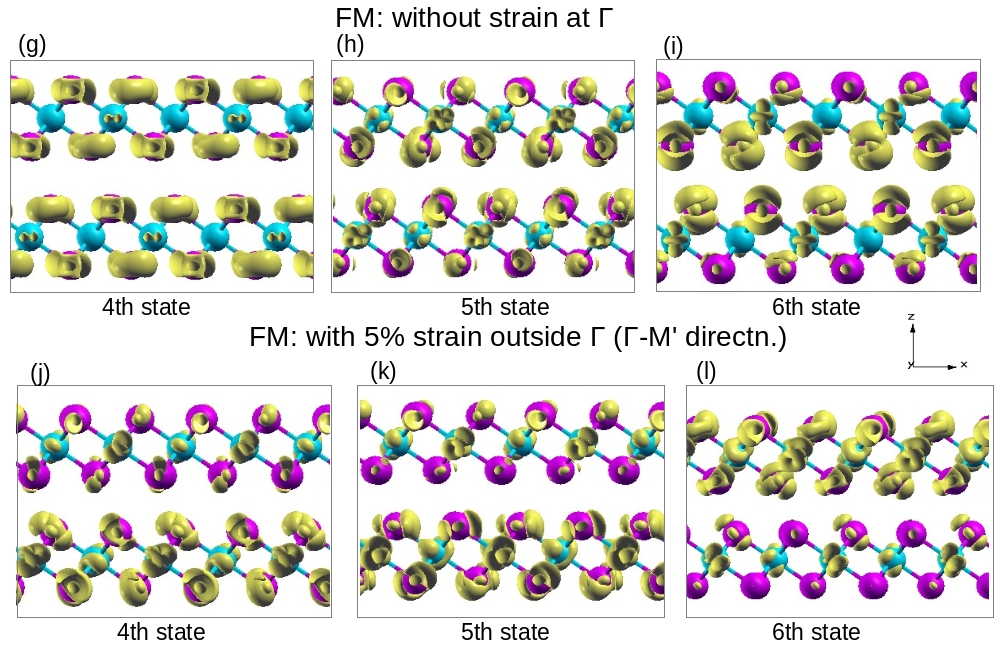}
\renewcommand{\thefigure}{S\arabic{figure}}
\caption{\label{3D_FM_second} 3D iso-surface plots of the probability density for the upper valence states of CrI$_{3}$ bilayer in FM configuration. (g)--(i) Unstrained bilayer at $\Gamma$, and (j)--(l) strained (5\%) bilayer outside $\Gamma$ along $\Gamma-M'$ direction. (g) and (j): Fourth upper valence state; (h) and (k): fifth upper valence state; (i) and (l): sixth upper  valence state for the unstrained and strained CrI$_{3}$ bilayers, respectively. Drawn at iso-surface value = 0.001 e/Bohr$^{3}$.}
\end{center}
\end{figure}

\begin{figure}[h]
\begin{center}
\includegraphics[width=0.95\textwidth]{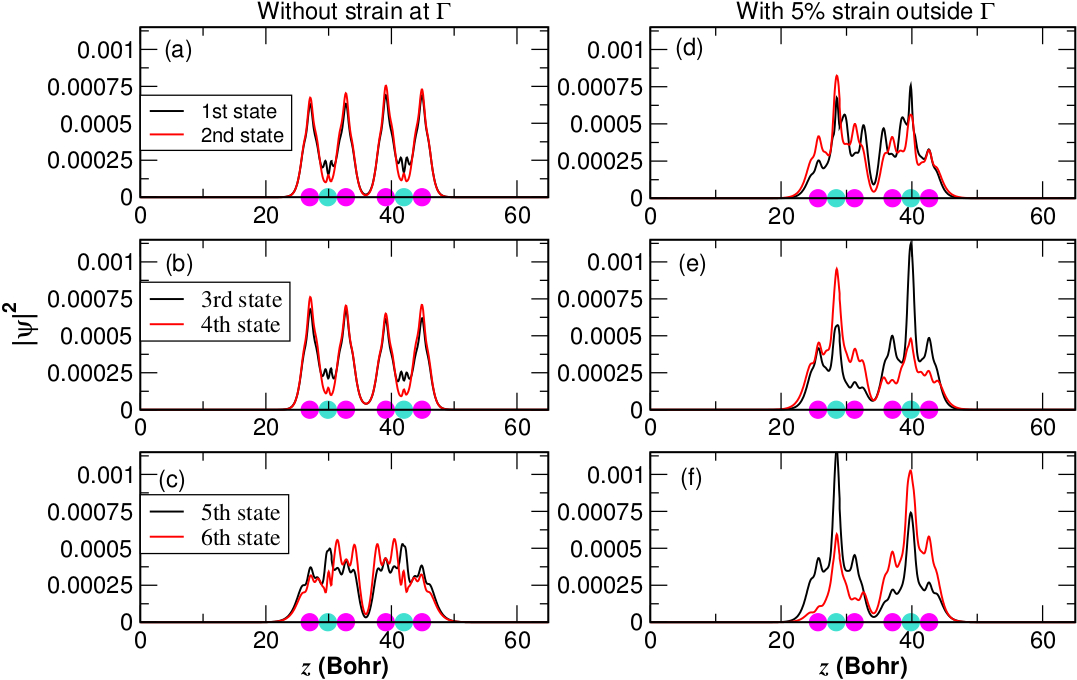}
\renewcommand{\thefigure}{S\arabic{figure}}
\caption{\label{avg_FM}Planar average plots of the probability density for the upper six valence bands of the pristine and strained CrI$_{3}$ bilayer in FM configuration. (a)--(c) Unstrained bilayer at $\Gamma$ and (d)--(f) strained bilayer outside $\Gamma$ (along $\Gamma-M'$ direction). (a) and (d) First and second upper valence states; (b) and (e) third and fourth upper valence states; (c) and (f) fifth and six upper valence states for the unstained and strained systems, respectively. The magenta and turquoise spheres show the iodine and chromium atoms, respectively.   }
\end{center}
\end{figure}

In figures~\ref{3D_FM_first} and ~\ref{3D_FM_second} we show the 3D isosurface plots for the probability density of the upper three valence bands in the pristine (at $\Gamma$) and strained FM configuration (near $\Gamma$, in the $\Gamma M'$ direction). The corresponding planar average plots of the probability densities are shown in figure~\ref{avg_FM}. 
Similar to the AFM case, in the pristine bilayer the highest four states have dominant I in-plane character (figures~\ref{3D_FM_first}, ~\ref{3D_FM_second} and  \ref{avg_FM}), while the next two valence states have strong out-of-plane probability density, in particular on the I atoms of the inner I planes of the bilayer. For the strained bilayer, the highest four states outside $\Gamma$ have pronounced out-of-plane features. These states, and particularly the upper two states, resemble the fifth and sixth upper valence states at $\Gamma$ of the unstrained FM bilayer. 

Hence, both for the FM and AFM bilayers, the pristine systems show a VBM  characterized by in-plane I states, while the application of the compressive strain induces a new local VBM for which the states display significant out-of-plane orbitals on the  atoms of the I inner planes of the  bilayer.


\end{document}